\documentclass[useAMS,usenatbib]{mn2e}

\usepackage{amsmath}
\usepackage{amssymb}
\usepackage[section]{placeins}
\usepackage{color}
\usepackage[usenames,dvipsnames]{xcolor}
\usepackage{mathrsfs}
\usepackage{mathtools}
\usepackage{natbib}
\usepackage{subfigure}
\usepackage{psfrag}
\usepackage{layouts}
\usepackage{upgreek}
\usepackage{graphicx}
\usepackage{soul}

\def\aap{A\&A}                
\def\aapr{A\&A~Rev.}          
\def\apj{ApJ}                 
\def\apjl{ApJ}                
\def\pasp{PASP}
\def\mnras{MNRAS}%
\def\pasj{PASJ}

\title[The pseudo-photosphere model for the continuum emission of gaseous disks]{The pseudo-photosphere model for the continuum emission of gaseous disks}
\author[R. G. Vieira, A. C. Carciofi and J. E. Bjorkman]{R. G. Vieira$^{1}$\thanks{E-mail:
rg.vieira@gmail.com}, A. C. Carciofi$^{1}$ and J. E. Bjorkman$^{2}$\\
$^{1}$Instituto de Astronomia, Geof\'isica e Ci\^encias Atmosf\'ericas,
      Universidade de S\~ao Paulo, Rua do Mat\~ao 1226, Cidade Universit\'aria,\\
      05508-900 S\~ao Paulo, SP, Brazil\\
$^{2}$Ritter Observatory, Department of Physics \& Astronomy, University of Toledo, Toledo, OH 43606, USA}
\begin{document}

\date{Accepted 2015 September 4.  Received 2015 September 3; in original form 2015 April 29}

\pagerange{\pageref{firstpage}--\pageref{lastpage}} \pubyear{2015}

\maketitle

\label{firstpage}


\begin{abstract}
The continuum emission of viscous decretion disks around Be stars is investigated. The results obtained from non-LTE radiative transfer models show two regimes in the disk surface brightness profile: an inner optically thick region, which behaves as a pseudo-photosphere with a wavelength-dependent size, and an optically thin tenuous outer part, which contributes with about a third of the total flux. The isophotal shape of the surface brightness is well described by elliptical contours with an axial ratio $b/a=\cos i$ for inclinations $i<75^{\circ}$. Based on these properties, a semi-analytical model was developed to describe the continuum emission of gaseous disks. It provides fluxes and spectral slopes at the infrared within an accuracy of $10\%$ and $5\%$, respectively, when compared to the numerical results. The model indicates that the infrared spectral slope is mainly determined by both the density radial slope and the disk flaring exponent, being practically independent of disk inclination and base density. As a first application, the density structure of 15 Be stars was investigated, based on the infrared flux excess, and the results compared to previous determinations in the literature. Our results indicate that the decretion rates are in the range of $10^{-12}$ to $10^{-9}\,{\rm M_{\odot}\,yr^{-1}}$, which is at least two orders of magnitude smaller than the previous outflowing disk model predictions.
\end{abstract}



\section{Introduction}
\label{introduction}

The conspicuous nature of classical Be stars was first noticed by the unexpected detection of an emission line in a stellar spectrum \citep{secchi1866}. Only a century later, the development of ground-based infrared (IR) observations allowed the detection of near-IR excess in Be stars \citep{johnson1967}. Although promptly related to the presence of circumstellar material, this information was not sufficient to distinguish between free-free and dust emission \citep{slettebak1988}. The emission origin was only decided by \citet{gehrz1974}, who interpreted the mid-IR excess of a sample of 33 Be stars as arising unambiguously from free-free emission. The gaseous nature of the circumstellar material was then confirmed, and the presence of dust ruled out. Soon after, \citet{wright1975} and \citet{panagia1975} simultaneously formulated the description of the free-free emission of expanding spherical envelopes. Based on this model, they were able to describe the observed spectral slope and estimate mass loss rates.

Almost a decade later, \citet{lamers1984} reformulated the theory of continuum emission from stellar winds, by introducing the curve of growth method. Among other contributions, these authors included the bound-free opacity in their models, showing that it is particularly important to explain the emission shortwards $10\,{\rm {\mu}m}$. Following the evidences of a disk-like structure at the time (e.g., \citealt{hartmann1978}), \citet{waters1986} introduced a wedge-shaped disk model with a given opening angle \citep{bjorkman1984}, and restricted the model calculations to pole-on orientations. Based on this model, \citeauthor{waters1987}~(1987, hereafter W87) interpreted IRAS observations of $101$ Be stars. Under the assumption of an outflowing disk with density structure of the form $\rho\propto r^{-n}$, they constrained the power law exponent to the interval $2<n<3.5$. These results are still widely quoted in the literature to this day.

In the last decade, the viscous decretion disk model (VDD; \citealt{lee1991}) became the new paradigm for Be star disks \citep{carciofi2011,rivinius2013}. It currently provides the better description to the diverse observational constraints available, and therefore has completely replaced the former outflowing disk models. Nevertheless, many authors still refer to the IR excesses, density structure and mass-loss constraints presented by W87. One of the goals of this study is to revisit the \citeauthor{waters1986}'s~(\citeyear{waters1986}) model, in the light of the currently accepted model for Be stars.

We introduce a simple, yet quite realistic, semi-analytical model to describe the continuum emission of a gaseous disk which is based on the assumption (verified by detailed numerical calculations) that their inner part behaves like a pseudo-photosphere. The term pseudo-photosphere is commonly used in the context of radiative transfer in supernovae (e.g., \citealt{bongard2008}), and was formerly introduced by \citet{leitherer1985} in the study of S~Dor variables. Such objects have temperature variations at constant luminosity, which were first interpreted as a consequence of a variable mass loss rate. An increase in the mass loss rate would eventually lead to the formation of an optically thick wind, originating an extended and cooler pseudo-photosphere (e.g., \citealt{davidson1987}). However, detailed calculations made by \citet{leitherer1989} and \citet{dekoter1996} have shown that S~Dor variations originate at subphotospheric regions. For a complete review on the subject, we refer to Section~4.2.2 of \citet{puls2008}. In the present context, we borrow the term to designate the optically thick region of a Be disk. Its formal definition and properties will be explored in the text. In the next section, we present the study of the disk brightness profile, computed with realistic radiative transfer (RT) simulations. The results obtained from this study motivated our semi-analytical model assumptions, which are described and applied to the proposed formulation (Sect.~\ref{model}). In Sect.~\ref{comparison}, we use the numerical results to validate the derived expressions, and some interesting implications of the model are discussed in Sect.~\ref{properties}. As a first application, we fit the IRAS data of a sub-sample from W87, and compare the results obtained from both models. The conclusions follow.


\section{The disk brightness profile}
\label{brightprofile}

The use of realistic RT codes has allowed not only the investigation of general properties of Be star disks (e.g., \citealt{carciofi2008}, \citealt{halonen2013}), but also the successful modeling of individual objects (e.g., \citealt{carciofi_etal2006,jones2008,carciofi2009}). One important result that emerges from these studies, and others that focused on the properties of large samples rather than individual stars (e.g., \citealt{touhami2011,silaj2014,meilland2012}), is the identification of kinematic viscosity as the main driver of Be disk outflows. In spite of all contributions brought by these studies, little attention has been devoted to one aspect: the disk continuum brightness profile, which bears important consequences not only to integrated quantities, such as flux, but most importantly to angularly resolved observations, such as near-IR interferometry.

To study the brightness profile of VDDs, we used {\ttfamily HDUST} \citep{carciofi2006}, a three-dimensional Monte Carlo RT code, capable of simultaneously calculating the non-LTE hydrogen level populations, ionization fraction and electron temperature from the radiative equilibrium at each disk position. We adopted a simplified parametric description of the VDD (e.g., \citealt{bjorkman2005})
\begin{equation}\label{rho}
\rho(\varpi,z)=\rho_0\,\left(\frac{\varpi}{R_{\star}}\right)^{-n}\exp\left(\frac{z^2}{2\,H^2} \right),
\end{equation}
where $\varpi$ and $z$ are respectively the radial and vertical cylindrical coordinates in the stellar frame of reference, and $R_{\star}$ is the stellar radius. The scale height is defined by the radial power law
\begin{equation}\label{height}
H(\varpi)=H_0\,\left(\frac{\varpi}{R_{\star}}\right)^{\beta},
\end{equation}
where $H_0=(c_s/V_{\textrm{crit}})\,R_{\star}$, $c_s=(kT_{\textrm{d}}/\mu m_{\textrm{H}})^{1/2}$ is the isothermal sound velocity, $V_{\textrm{crit}}=(GM_{\star}/R_{\star})^{1/2}$ is the break-up velocity, $M_{\star}$ is the stellar mass, $T_{\textrm{d}}$ is the disk temperature, $\mu$ is the mean molecular weight of the gas, $m_H$ is the proton mass, $k$ is the Boltzmann constant and $\beta$ is a free parameter to describe the disk flaring. The physical motivation for adopting a simple power law for the density is as follows. For isothermal Be disks fed at a (nearly) constant rate from the central star, it can be demonstrated that the radial density is a power law with $n=3.5$. However, disks either subjected to varying mass feeding rates, or undergoing build-up or dissipation phases, will have much more complex density slopes \citep{haubois2012}, which are here simulated by allowing $n$ to be a free parameter.

With the purpose of developing a systematic analysis, a small set of representative {\ttfamily HDUST} models were computed. The disk model was chosen to be isothermal, to allow the comparison between the {\ttfamily HDUST} results and the semi-analytical model (Sect.~\ref{comparison}). Following \citet{carciofi2006}, the adopted disk temperature was $T_{\textrm{d}}=0.6\,T_{\textrm{\textrm{eff}}}$, where $T_{\textrm{eff}}$ is the stellar effective temperature.

The stellar parameters were chosen to be compatible with the evolutionary stellar models of \cite{georgy2013}. For a set of fixed values for mass, rotation rate and age, the other stellar parameters are determined by the evolutionary models. We chose a non-rotating stellar model, since the rotation effects are out of the scope of the present work. Such effects are left to forthcoming publications.

The adopted stellar parameters are described in Table~\ref{stellar_par}, while the chosen disk parameters are listed in Table~\ref{disk_par}. For each model, the spectrum and corresponding synthetic images were computed from $3500\,{\rm \AA}$ to $100\,{\rm {\mu}m}$. Finally, the disk size $R_{\textrm{d}}=50\,R_{\star}$ was chosen to be sufficiently large to not affect our results at longer wavelengths.


\subsection{The pseudo-photosphere}
\label{dual}

\begin{figure}
\begin{center}
\includegraphics[angle=0,scale=.8]{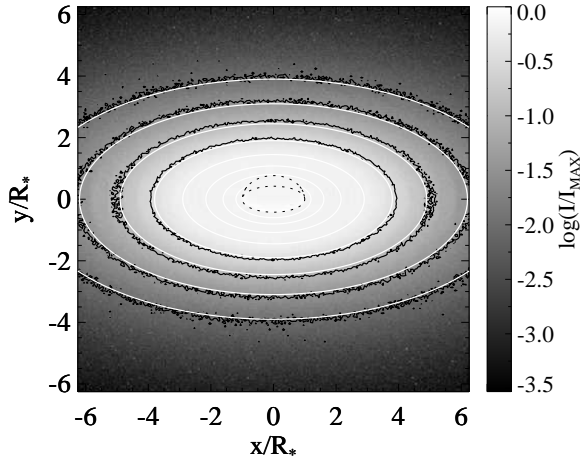}
\caption{Synthetic image at $\lambda=10\,{\rm {\mu}m}$ computed with {\ttfamily HDUST}, normalized by its maximum value. The model parameters are $\rho=2.8\times 10^{-11}\,{\rm g\,cm^{-3}}$, $n=3.5$, $i=60^{\circ}$ and B1 spectral subtype. The image isophotal contours are shown (black), while white ellipses define the adopted annuli, used in the calculation of the disk brightness profile. The stellar boundaries are also indicated in the figure (dotted line) \label{annuli}}
\end{center}
\end{figure}

\begin{figure}
\begin{center}
\includegraphics[angle=0,scale=.7]{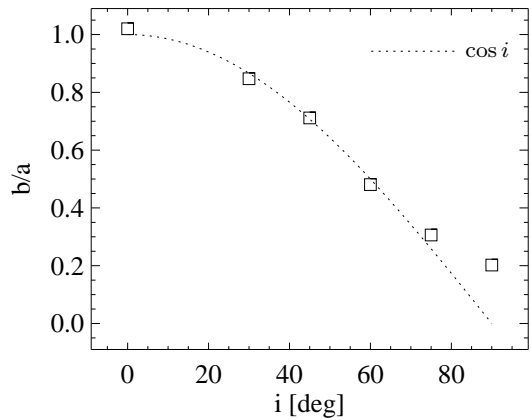}
\caption{Ellipse parameters obtained from the isophotal contour fit at $10\,{\rm R_{\star}}$, for several disk inclinations and same model parameters as for Fig.~\ref{annuli}. $a$ and $b$ are the semi-major and semi-minor axes of the isophotes, respectively. The dotted curve shows the expected result for geometrically thin disks. \label{ellipse_fit}}
\end{center}
\end{figure}

Brightness profile curves were computed from the synthetic images. Using a logarithmically spaced radial grid, we integrated the flux over isophotal elliptical annuli (excluding the star). Figure~\ref{annuli} shows the adopted grid, superimposed on the central region of a sample synthetic image. Since the stellar surface blocks a portion of the inner disk, the flux arising from this region was estimated by integrating the un-shadowed disk flux (bottom half of the annulus) and then multiplying it by two. In order to verify if the choice of elliptical annuli is adequate, we fitted ellipses to the isophotes at $10\,{\rm R_{\star}}$ of a representative model, for several inclinations. The results are presented in Fig.~\ref{ellipse_fit}, where we compare the ratio of the isophotal axes to the expectation of a perfect ellipse. As expected, the approximation is satisfactory for non edge-on cases, and starts to fail at $i \gtrsim 75^{\circ}$.

\begin{table}
\caption{List of adopted stellar parameters.}
\label{stellar_par}
\begin{center}
\begin{tabular}{lcc}
\hline\hline
Parameter	&	\multicolumn{2}{c}{Value}	\\	\hline
Sp. Type	&	B1V	&	B3V	\\
$M_{\star}/{\rm M_{\odot}}\,^a$	&	$12.5$	&	$7.7$	\\
$L_{\star}/{\rm L_{\odot}}$	&	$17800$	&	$3660$	\\
$R_{\star}/{\rm R_{\odot}}$	&	$6.57$	&	$4.94$	\\
$T_{\textrm{eff}}$	&	$26000$~K	&	$20200$~K	\\
$t/t_{\textrm{MS}}$	&	\multicolumn{2}{c}{$0.75$}	\\
\hline
\end{tabular}
\end{center}
(a) \citet{schmidt1982}.
\end{table}

\begin{table}
\caption{List of adopted disk parameters.}
\label{disk_par}
\begin{center}
\begin{tabular}{lccc}
\hline\hline
Parameter	&	\multicolumn{3}{c}{Value}	\\ \hline
$\rho_0/{\rm [g\, cm^{-3}]}$	&	$8.4\times 10^{-12}$	&	$2.8\times 10^{-11}$	&	$8.4\times 10^{-11}$ \\
$n$	&	$3.0$	&	$3.5$	&	$4.0$	\\
$i$	&	\multicolumn{3}{c}{$0^{\circ}$ \quad  $30^{\circ}$ \quad  $45^{\circ}$ \quad $60^{\circ}$ \quad  $75^{\circ}$ \quad  $90^{\circ}$}	\\
$T_{\textrm{d}}/T_{\textrm{eff}}$	&	\multicolumn{3}{c}{$0.6$}	\\
$\beta$	&	\multicolumn{3}{c}{$1.5$}	\\
$ R_{\textrm{d}}/{\rm R_{\star}}$	&	\multicolumn{3}{c}{$50$}	\\
\hline
\end{tabular}
\end{center}
\end{table}

\begin{figure*}
\begin{center}
\includegraphics[angle=0,scale=.8]{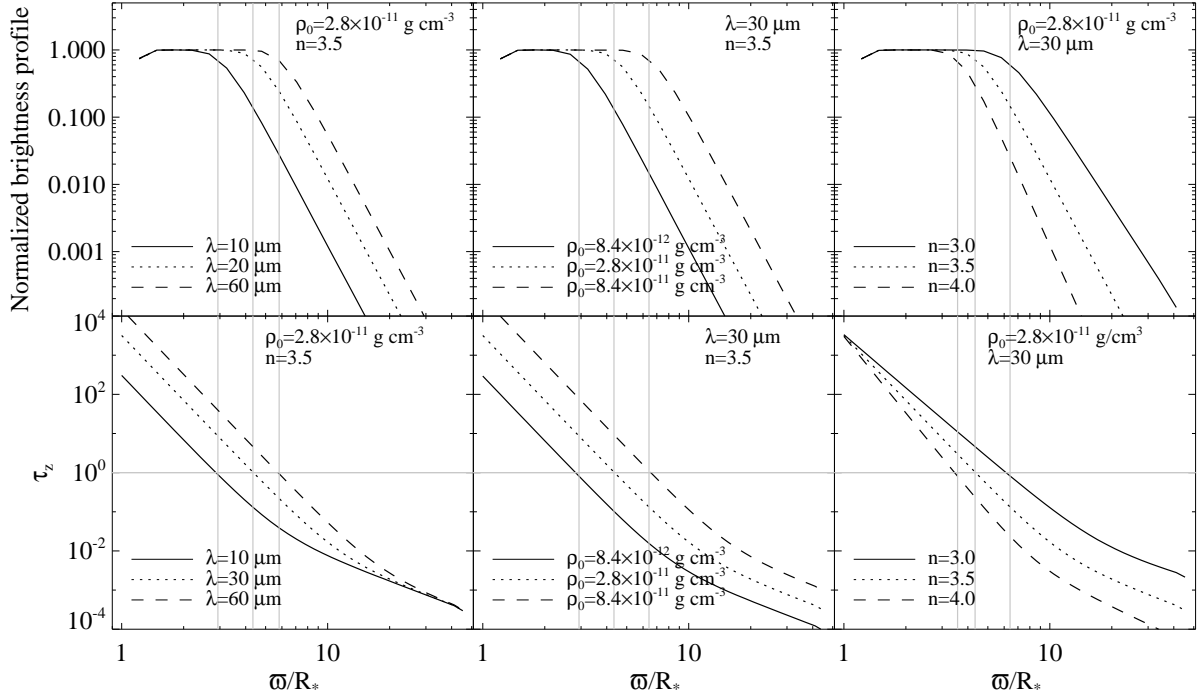}
\caption{Brightness profiles of representative models computed with {\ttfamily HDUST} (upper panels), and respective vertical optical depth radial dependence (bottom panels). The models have pole-on orientation and B1 spectral subtype. The vertical dotted lines indicate the position where $\tau_z=1$. \label{bright_profile}}
\end{center}
\end{figure*}

The pole-on brightness profile for a representative set of models is shown in the top row of Fig.~\ref{bright_profile}. Note that the little dip in brightness close to the stellar surface is just an artifact stemming from the adopted radial spacing, which becomes smaller than the image pixel size at the inner disk. The radial dependence of the vertical continuum optical depth is shown in the bottom row. The results for the B3 models are very similar. There are two regimes in the brightness profiles, which are clearly controlled by the vertical optical depth. The (vertically) optically thick inner part is responsible for the bulk of the disk emission, and has a very flat radial dependence. The optically thin outer region is characterized by a weaker emission, which falls sharply with radius. Not surprisingly, the transition region between these two regimes always occurs close to the position where $\tau_z\simeq 1$.

\begin{figure}
\begin{center}
\includegraphics[angle=0,scale=.7]{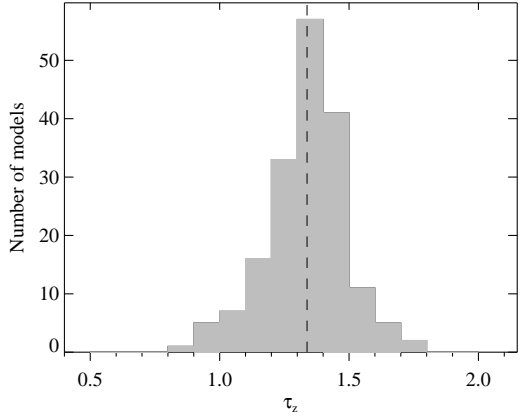}
\caption{Vertical optical depth $\tau_z(\lambda,\overline{R})$ distribution. The effective radius values were estimated by the fit of the disk brightness profile with a broken power law, at 10 wavelengths covering the adopted spectral range. The dashed line represent the mean of the $\tau_z$ distribution. To ensure a reliable power law fit, only the models with at least five radial positions where $\tau_z>1$ were included. \label{tauz_ref}}
\end{center}
\end{figure}

Let us now investigate more closely the relation between the size of the optically thick inner part and the vertical optical depth. To quantify this, we define an effective radius, $\overline{R}$, by fitting the brightness profile with a broken power law and setting $\overline{R}$ as the point where the slope changes. The relation between the derived transition positions and the respective vertical optical depths is presented in Fig.~\ref{tauz_ref}, again only for the pole-on models. The $\tau_z$ mean value is indeed close to unity ($\langle\tau_z\rangle\approx 1.3$), but the distribution has a considerable dispersion, which may be partly attributed to the fact that the transition region is actually smoother than a broken power law.

\psfrag{cosi}[c][][1.5]{$\cos i$}

The optically thick inner part of the disk may be regarded as a pseudo-photosphere with an effective radius, $\overline{R}$, that marks the discontinuity in slope of the brightness distribution. We define $\overline{R}=\overline{R}(\lambda,i,\rho_0,n,\beta;M_{\star},R_{\star},T_{\textrm{d}})$ as the position of the disk for which the optical depth in the line of sight is close to the unity. Note that our definition for $\overline{R}$ differs from the one proposed by \cite{wright1975}, defined for a spherical envelope at $\tau_{\textrm{radial}}=0.244$.

\begin{figure}
\begin{center}
\includegraphics[angle=0,scale=.7]{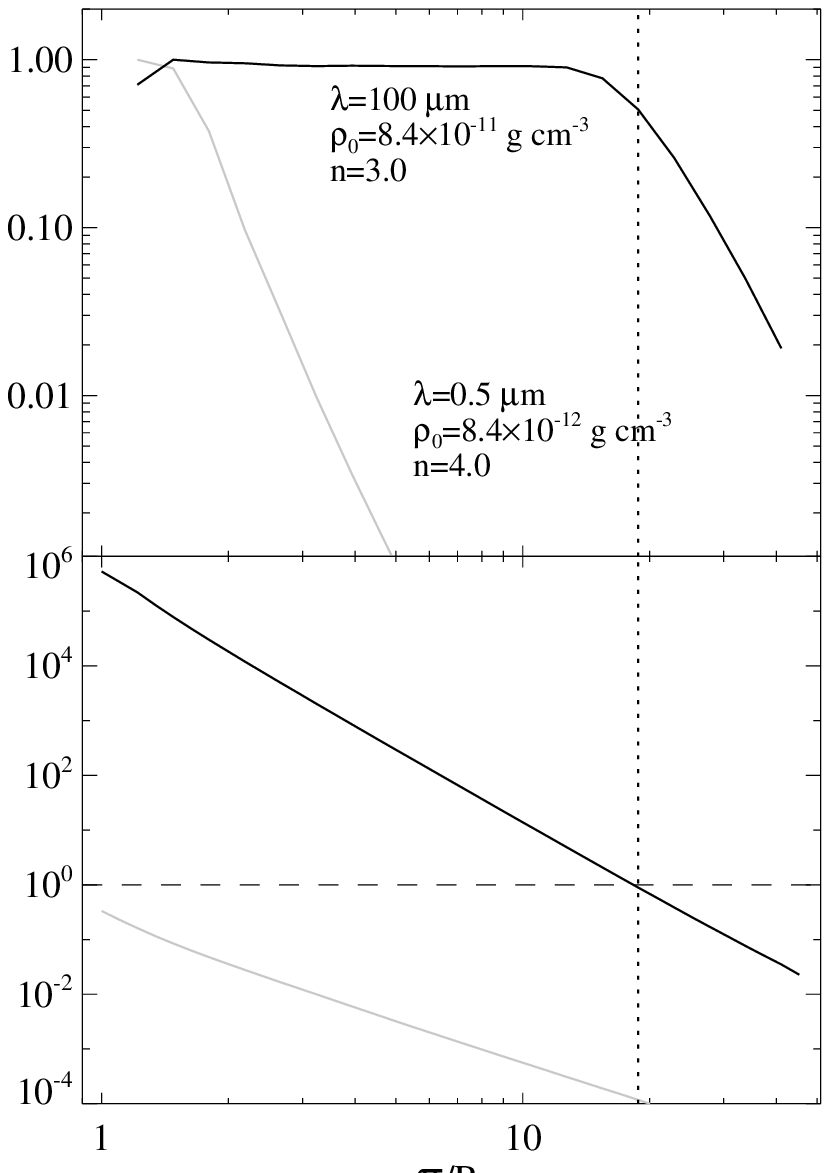}
\caption{Special cases where {\it (i)} pseudo-photosphere has practically the disk size (solid line) and {\it (ii)} pseudo-photosphere is absent (dotted line). The vertical line represents the position where $\tau_z=1$ for the first model. Both models have B3 spectral subtype and pole on orientation.\label{bright_extreme}}
\end{center}
\end{figure}

Given that the two regimes of the disk emission are controlled by the vertical optical depth, one can envisage two situations for which only one of these regimes is present. They are illustrated in Fig.~\ref{bright_extreme} for the case where $\overline{R}$ is of the order of $R_{\textrm{d}}$, and for the case where the brightness profile slope discontinuity is absent (i.e., there is no pseudo-photosphere). In both cases, one component (optically thick or thin) dominates the disk emission.

The study of the disk brightness profile revealed some general properties of the disk continuum emission. In the next section, a simple, yet realistic, analytical model is developed in order to relate the system physical parameters to the disk brightness distribution.

\section{The pseudo-photosphere disk model}
\label{model}


\subsection{Effective radius}
\label{effective_radius}

As we shall demonstrate, the most fundamental parameter to describe the continuum emission of a gaseous disk at a given wavelength is the size of the pseudo-photosphere (effective radius) at that wavelength. In order to compute its value from first principles, we adopt the following opacity expression \citep{brussaard1962}:
\begin{align}\label{kappa}
\kappa_{\lambda} &= 3.692\times 10^8\, \left[1-\exp\left(-hc/\lambda k T_{\textrm{d}}\right)\right]\overline{z^2} T_{\textrm{d}}^{-1/2} \nonumber \\
                 & \quad\times\left(\lambda/c\right)^{3} \gamma \left(\rho/\mu\,m_{\textrm{H}}\right)^2 \left[g(\lambda,T_{\textrm{d}})+b(\lambda,T_{\textrm{d}})\right],
\end{align}
where $\overline{z^2}$ is the mean value of the square atomic number, $\gamma$ is the ionization fraction, and the quantities inside the brackets are the free-free and bound-free gaunt factors, respectively. For simplicity, we assume that the disk is isothermal, and express its temperature as a fraction of the stellar one, written as $T_{\textrm{d}}=f\,T_{\textrm{eff}}$. For a discussion of the non-isothermal case, we refer to \cite{vieira2014}. Using Eqs.~(\ref{rho}) and (\ref{height}), the vertical optical depth at a given radius is then given by
\begin{equation}\label{tauz_eq}
\tau_z(\lambda,\varpi)=\int_{-\infty}^{+\infty} \kappa_{\lambda}(\varpi,z)\, dz=\tau_0\, \left(\frac{\varpi}{R_{\star}}\right)^{-2n+\beta}
\end{equation}
where
\begin{equation}\label{tau0}
\tau_0=\sqrt{\pi}\,H_0\,\kappa_{\lambda}(\varpi=R_{\star},z=0).
\end{equation}

Note that we implicitly assumed an axi-symmetric disk, and thus $\kappa_{\lambda}$ is independent of azimuth. For an observer along the direction having an angle $i$ with respect to the disk rotation axis, the optical depth along the line of sight is
\begin{equation}\label{tau_inclination}
\tau_i(\lambda,\varpi')=\sec i\,\tau_z=\tau_0\,\sec i\,\left(\frac{\varpi'}{R_{\star}}\right)^{-2n+\beta},
\end{equation}
where $\varpi'$ the radial distance along the semi-major axis. Since the disk projection on the sky has an elliptical shape, each $\varpi'$ position specifies the optical depth for the entire corresponding elliptical annulus (except the disk area shadowed by the star). This inclination dependence holds only if the opacity does not change along the optical path, which is approximately valid for a geometrically thin disk. Note that by adopting this expression, we have implicitly excluded the edge-on case from our discussion. A comparison between the analytical formulae here derived with numerical results (Sect.~\ref{comparison}) indicates that this approximation holds for $i \lesssim 75^{\circ}$. By defining the effective radius as
\begin{equation}\label{ref_def}
\tau_i(\lambda,\overline{R})=\overline{\tau},
\end{equation}
where $\overline{\tau}$ is a free parameter of the order of unity, we have that
\begin{equation}\label{ref_eq}
\frac{\overline{R}}{R_{\star}}=\left(\sec i\,\frac{\tau_0}{\overline{\tau}} \right)^{1/(2n-\beta)}.
\end{equation}
$\overline{R}$ is a function of wavelength, the stellar/disk physical parameters (via $\tau_0$), and the direction $i$ through which it is observed. In particular, the wavelength dependence is very strong, and goes as
\begin{equation}
\overline{R}\propto\lambda^{(2+u)/(2n-\beta)},
\end{equation}
where
\begin{equation}
u=\frac{d\ln(g+b)}{d\ln\lambda}.
\end{equation}


\subsection{Disk emission}
\label{disk_emission}

The specific intensity of the star plus an isothermal disk can be written as
\begin{equation}\label{bright_general}
I_{\lambda}(\varpi')=
\begin{cases}
B_{\lambda}(T_{\textrm{eff}}) & \textrm{($A_{\star}^{1/2}$)}\\
B_{\lambda}(T_{\textrm{eff}})\,e^{-\tau_i}+B_{\lambda}(T_{\textrm{d}})\,[1-e^{-\tau_i}] & \textrm{($A_{\star}^{-1/2}$)} \\
B_{\lambda}(T_{\textrm{d}})\,[1-e^{-\tau_i}] & \textrm{($A_{\textrm{disk}}$),}\\
\end{cases}
\end{equation}
where $A_{\star}^{1/2}$ and $A_{\star}^{-1/2}$ represent the areas of the stellar upper and lower (hidden by the disk) hemispheres respectively, and $A_{\textrm{disk}}$ represents the visible disk surface (see Fig.~\ref{cases}). For simplicity, we describe the photospheric emission by a black body. The effects of limb-darkening, stellar rotation and circumstellar extinction are neglected. The disk source function was assumed to be constant along the line of sight, which is a good approximation for a non edge-on geometrically thin disk. In the following, we consider the general case (Sect.~\ref{dual_disk}), followed by the special cases of a tenuous disk (Sect.~\ref{diff_disk}) and a truncated pseudo-photosphere (Sect.~\ref{trunc_disk}).


\subsubsection{General case}
\label{dual_disk}

\begin{figure}
\begin{center}
\includegraphics[angle=0,scale=0.4]{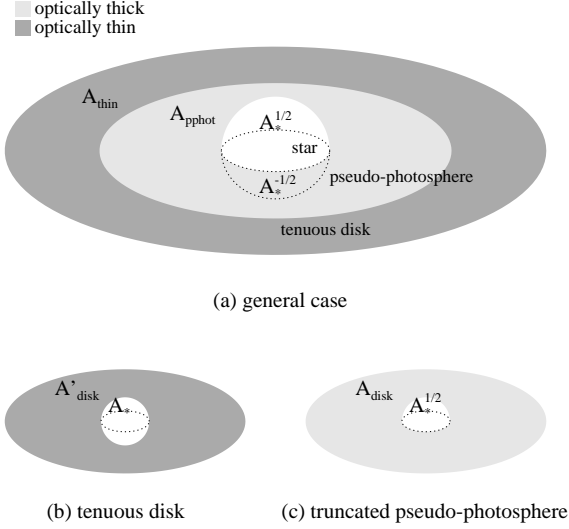}
\caption{Schematic representation of disk components, for the three possible cases: (a) both emission regimes are present, where $R_{\star}<\overline{R}<R_{\textrm{d}}$; (b) optically thin case, where $\overline{R}\leq R_{\star}$; and (c) truncated pseudo-photosphere, where $\overline{R}\leq R_{\textrm{d}}$. \label{cases}}
\end{center}
\end{figure}

Following Sect.~\ref{dual}, the disk emission is here approximated by an optically thick inner region ($\tau_i>1$) and an optically thin outer part ($\tau_i<1$), separated by the effective radius (Fig.~\ref{cases}a). Under this assumption, Eq.~(\ref{bright_general}) can be rewritten as
\begin{equation}\label{bright_dual_eq}
I_{\lambda}(\varpi')= B_{\lambda}(T_{\textrm{eff}})
\begin{cases}
1 & \textrm{($A_{\star}^{1/2}$)}\\
\mathscr{F} & \textrm{($A_{\textrm{pphot}}$)}\\
\overline{\tau}\mathscr{F}\, (\varpi'/\overline{R})^{-2n+\beta} & \textrm{($A_{\textrm{thin}}$),}\\
\end{cases}
\end{equation}
where $A_{\textrm{pphot}}$  and $A_{\textrm{thin}}$ represent respectively the pseudo-photosphere and optically thin region visible surfaces, and we use the convenient notation
\begin{equation}
\mathscr{F}\equiv\frac{B_{\lambda}(T_{\textrm{d}})}{B_{\lambda}(T_{\textrm{eff}})}.
\end{equation}
These simplifications allow us to derive an analytical expression for the integrated flux
\begin{align}\label{flux_dual}
F_{\lambda} &= F_{\lambda}^{\star}\left\{\cos i\left[ \left(\frac{2n-\beta
+\delta}{2n-\beta-2} \right)\left(\frac{\overline{R}}{R_{\star}}\right)^2\mathscr{F}+\frac{1}{2}\left(1-\mathscr{F}\right)\right] \right. \nonumber\\
                            & \quad \left. +\frac{1}{2}\left(1-\mathscr{F}\right) \right\},
\end{align}
where the stellar flux is given by
\begin{equation}\label{fstar}
F_{\lambda}^{\star}=\pi B_{\lambda}(T_{\textrm{eff}})\,\left(\frac{R_{\star}}{d} \right)^2,
\end{equation}
$d$ is the distance to the star, and
\begin{equation}
\delta=2\overline{\tau}\left[1-(\overline{R}/R_{\textrm{d}})^{2n-\beta-2}\right]-2.
\end{equation}
Note that $\delta\rightarrow 0$ if $\overline{R}\ll R_{\textrm{d}}$ and $\overline{\tau}=1$. Finally, it is easy to rewrite the equations in the Rayleigh-Jeans regime, by making use of the relation
\begin{equation}
\lim_{\lambda \to \infty} \mathscr{F}=f.
\end{equation}

We can also interpret the SED slope (or, equivalently, broad-band colors) based on the derived flux expressions. First, we define the spectral slope as
\begin{equation}\label{alpha_def}
\alpha_{\textrm{IR}}\equiv-\frac{d\ln F_{\lambda}}{d\ln\lambda}.
\end{equation}
Although straightforward, the logarithmic differentiation of Eq.~(\ref{flux_dual}) leads to very complicated expressions. In order to derive simpler analytical formulae for $\alpha_{\textrm{IR}}$, we assume the Rayleigh-Jeans limit for $B_{\lambda}(T_{\textrm{eff}})$. For the case where $R_{\star}<\overline{R}<R_{\textrm{d}}$, we have
\begin{align}\label{slp_dual_eq}
\alpha_{\textrm{IR}} &=4-\frac{2\,f\,\cos i\,(2+u)}{Z_{\lambda}\,(2n-\beta)} \left(\frac{\overline{R}}{R_{\star}}\right)^2 \nonumber \frac{(2n-\beta-2+2\overline{\tau})}{(2n-\beta-2)}\\
                            & \quad\times\left[1-\left(\overline{R}/R_{\textrm{d}}\right)^{2n-\beta-2}\right]
\end{align}
where $Z_{\lambda}\equiv F_{\lambda}/F_{\lambda}^{\star}$. In particular, if we assume that $\overline{R}\ll R_{\textrm{d}}$ and consider the limit for large $\overline{R}$, the previous equation is simplified to
\begin{equation}\label{slp_lim_eq}
\lim_{\overline{R} \to \infty} \alpha_{\textrm{IR}}=4-\frac{4+2u}{2n-\beta},
\end{equation}
where we also assumed that $\overline{\tau}=1$. As we shall verify in Sect.~\ref{properties}, the dependence of the spectral slope on disk inclination and base density is very weak, and the large $\overline{R}$ limit is usually valid at the mid/far-IR.


\subsubsection{Tenuous disk}
\label{diff_disk}

The case where $\overline{R}\leq R_{\star}$ corresponds to the situation where the disk is entirely optically thin at the line of sight direction ($\tau_i<1$, Fig.~\ref{cases}b). In this case,
\begin{equation}\label{bright_diff_eq}
I_{\lambda}^{\textrm{thin}}(\varpi')=B_{\lambda}(T_{\textrm{eff}})
\begin{cases}
1 & \textrm{($A_{\star}$)}\\
\overline{\tau}\mathscr{F}\, (\varpi'/\overline{R})^{-2n+\beta} & \textrm{($A_{\textrm{disk}}'$),}\\
\end{cases}
\end{equation}
where $A_{\star}$ stands for the entire stellar surface (note the disk is optically thin so there is no relevant circumstellar extinction), and $A'_{\textrm{disk}}$ is the disk surface not intersecting $A_{\star}$ (Fig.~\ref{cases}b). Even though $\overline{R}$ no longer has a geometrical interpretation, it still appears in Eq.~(\ref{bright_diff_eq}) as an optical depth scale, as evidenced by rearranging Eq.~(\ref{ref_eq}) as
\begin{equation}
\tau_0=\overline{\tau}\,\cos i\,\left(\frac{\overline{R}}{R_{\star}}\right)^{2n-\beta}.
\end{equation}
The flux expression for this case is
\begin{equation}\label{flux_diff}
F_{\lambda}^{\textrm{thin}}=F_{\lambda}^{\star}\left[1+\frac{2\overline{\tau}\mathscr{F}\,\cos i}{2n-\beta-2}\left(\frac{\overline{R}}{R_{\star}}\right)^{2n-\beta}\right],
\end{equation}
where we assumed that $\overline{R}\ll R_{\textrm{d}}$, given that $\overline{R}\leq R_{\star}$ in this case. Similarly to the general case, we can derive the spectral slope expression for the optically thin case (also for $\overline{R}\leq R_{\star}$ and $\overline{R}\ll R_{\textrm{d}}$):
\begin{equation}\label{slp_diff_eq}
\alpha_{\textrm{IR}}^{\textrm{thin}}=4-\frac{2\overline{\tau}f\,\cos i\,(2+u)}{Z_{\lambda}(2n-\beta-2)} \left(\frac{\overline{R}}{R_{\star}}\right)^{2n-\beta},
\end{equation}
again in the Rayleigh-Jeans limit.


\subsubsection{Truncated pseudo-photosphere}
\label{trunc_disk}

The situation where $\overline{R}\geq R_{\textrm{d}}$ (Fig.~\ref{cases}c) typically happens when the disk is quite small (i.e., truncated by a binary companion) and/or at quite long wavelengths (e.g., in the sub-mm or radio domains). For this case, both the pseudo-photosphere and the disk have the same size, and the specific intensity is simply given by
\begin{equation}\label{bright_trunc_eq}
I_{\lambda}^{\textrm{trunc}}(\varpi')=B_{\lambda}(T_{\textrm{eff}})
\begin{cases}
1 & \textrm{($A_{\star}^{1/2}$)}\\
\mathscr{F} & \textrm{($A_{\textrm{disk}}$),}\\
\end{cases}
\end{equation}
with respective flux in the form
\begin{equation}\label{flux_trunc}
F_{\lambda}^{\textrm{trunc}} = F_{\lambda}^{\star}\left[\frac{1}{2}\,(1-\mathscr{F})\,(1+\cos i)+\mathscr{F}\,\cos i\,\left(\frac{R_{\textrm{d}}}{R_{\star}}\right)^2 \right].
\end{equation}
The only wavelength dependence in this solution comes from the stellar flux term, and therefore the observed spectral dependence should go as a simple black body. For the case where $\overline{R}\geq R_{\textrm{d}}$, and assuming the Rayleigh-Jeans regime, the spectral slope is simply
\begin{equation}\label{slp_trunc_eq}
\alpha_{\textrm{IR}}^{\textrm{trunc}}=4.
\end{equation}

Finally, it is important to note that the choice of the appropriate expression for the flux (Eqs.~\ref{flux_dual}, \ref{flux_diff} or \ref{flux_trunc}) depends on the effective radius, which in turn depends on wavelength. Consequently, a typical observed SED may result from all three situations: tenuous disk emission at optical/near-IR, the general case at mid/far-IR, and the truncated pseudo-photosphere case at radio wavelengths.

Despite the adopted simplifications, we demonstrate below that the above formulae are reasonably accurate for calculating the SED of gaseous disk, and can be successfully used to infer physical parameters from these disks from observations.


\section{Comparison with numerical simulations}
\label{comparison}

In order to evaluate Eq.~(\ref{ref_eq}) for comparison with the numerical models of Sect.~\ref{effective_radius}, we fixed the parameters $f=0.6$, $\mu=0.5$ (ionized hydrogen), $\overline{z^2}=1$ (hydrogen only), $\gamma=1$ (fully ionized disk) and $\overline{\tau}=1.3$. Additionally, we adopted the gaunt factors of \citet{hummer1988} and \citet{storey1991}. Useful approximations for the free-free and bound-free gaunt factors are presented in Appendix~\ref{gaunt_factor}. Also, we applied a correction factor to the black body stellar flux, in order to make it compatible with the stellar photospheric model used by {\ttfamily HDUST}. Such correction is described in Appendix~\ref{bb_corr_factor}. Fig.~\ref{ref_lbd} compares the spectral dependence of $\overline{R}$ given by the analytical model with the numerical results, for different $\rho_0$ and $n$ values. We observe very good agreement between the results obtained from the different methods. The effective radius dependence with the disk inclination is shown in Fig.~\ref{ref_inc} for a representative model. The satisfactory match between the models suggests that Eq.~(\ref{tau_inclination}) is a good approximation for the optical depth for $i\leq 75^{\circ}$.


\psfrag{refrs}[c][][1.7]{$\overline{R}/R_{\star}$}

\begin{figure}
\begin{center}
\includegraphics[angle=0,scale=0.6]{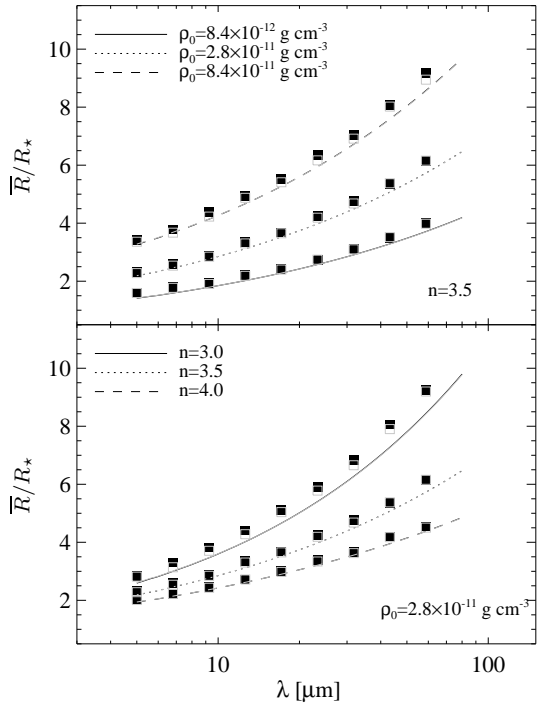}
\caption{Comparison between the $\overline{R}$ values estimated with the broken power law fit of the {\ttfamily HDUST} brightness profiles, and those obtained from the analytical model. The filled squares correspond to B1 spectral subtype, while the and open squares to B3. The gray lines representing the analytical model are practically superimposed, and represent respectively the B1 and B3 spectral subtypes. All the models were considered to be seen at pole-on orientation. \label{ref_lbd}}
\end{center}
\end{figure}

The flux excess predictions are compared in Fig.~\ref{ir_excess}. Typically, our results are accurate within $10\%$, when compared with the {\ttfamily HDUST} results. The discrepancies increase at the optical/near-IR wavelengths, where the bound-free opacity dominates the continuum optical depth, because {\ttfamily HDUST} self-consistently calculates the H level populations in the non-LTE regime, whereas Eq.~({\ref{kappa}) implicitly assumes LTE populations. Finally, as expected, the analytical formulae start to break down around $75^{\circ}$, where the assumption made in Eq.~(\ref{tau_inclination}) no longer holds.

The spectral slopes are compared in Fig.~\ref{ir_slope}. For $\lambda\gtrsim 3\,{\mu{m}}$, the analytical model predicts slopes that are typically within $5\%$ of the numerical results, and, as expected, the accuracy is better longwards of the mid-infrared.

The numerical results also agree with the analytical expectations that the spectral slope depends very little on the disk orientation, as opposed to the flux excess. This is because the flux level is directly proportional to the disk emitting area, which decreases with increasing inclination angle. On the other hand, the spectral slope is mainly determined by the disk density structure, according to Eq.~(\ref{slp_lim_eq}). Notably, the edge-on case represents an exception in this discussion. The stellar light extinction and disk self-absorption become very important at this orientation, thus requiring a different model approach that does not employ a thin disk approximation.


\psfrag{refrs}[c][][1.5]{$\overline{R}/R_{\star}$}

\begin{figure}
\begin{center}
\includegraphics[angle=0,scale=0.7]{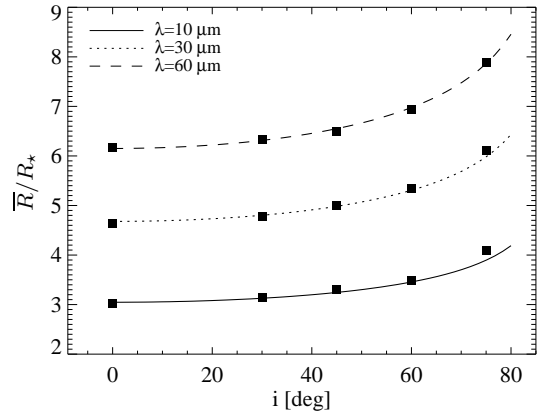}
\caption{Dependence of $\overline{R}$ with inclination at several wavelengths, as indicated. The chosen model has B1 spectral subtype, $\rho_0=2.8\times 10^{-11}\,{\rm g\,cm^{-3}}$ and $n=3.5$. \label{ref_inc}}
\end{center}
\end{figure}


\section{Properties of the solution}
\label{properties}


\subsection{Flux contribution from distinct disk regions}
\label{fractions}

Having demonstrated the validity of the pseudo-photosphere model in reproducing the disk continuum emission of gaseous disks, the properties of the solution are now examined. The description of the emission of the star and the disk in terms of distinct components (star, pseudo-photosphere, and tenuous disk) naturally raises the question of what are their relative contributions to the total flux at a given wavelength. The relative contributions of each component are shown in Fig.~\ref{flux_fraction} as a function of effective radius.

\begin{figure*}
  \begin{center}
        \subfigure[Flux excess.]{%
            \label{ir_excess}
            \includegraphics[width=.6\linewidth]{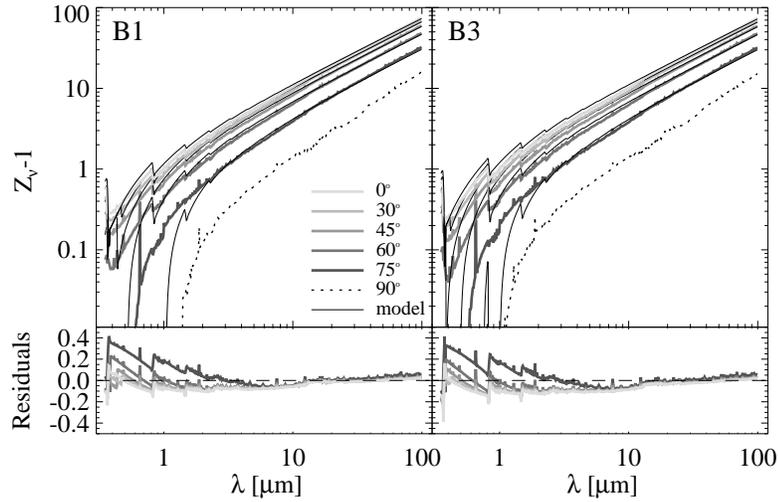}
        }\\
        \subfigure[Spectral slope.]{%
            \label{ir_slope}
            \includegraphics[width=.6\linewidth]{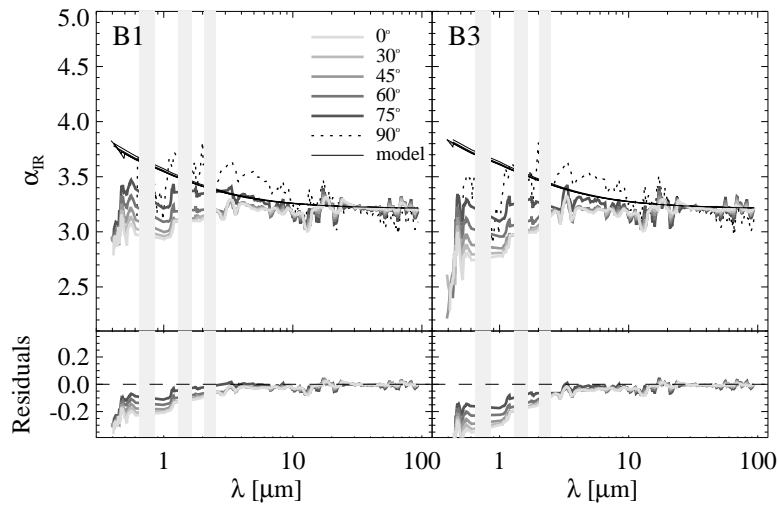}
        }\\
  \end{center}
  \caption{Comparison between the flux excess and spectral slope computed with {\ttfamily HDUST} and the pseudo-photosphere model, for a disk with $\rho_0=2.8\times 10^{-11}\,{\rm g\, cm^{-3}}$ and $n=3.5$. The bottom panels show the respective residuals. Each gray shade represents an inclination angle, as indicated, and the spectral slope discontinuities were excluded (vertical gray bars). The edge-on {\ttfamily HDUST} model is also presented, for reference, but no comparison is made with the analytical model in this case.}
  \label{ir_flux}
\end{figure*}

For very small $\overline{R}$, the stellar and tenuous disk emissions dominate the flux. The situation is rapidly inverted as $\overline{R}$ increases, and the stellar flux becomes negligible at sufficiently large $\overline{R}$ ($\gtrsim 5\,R_{\star}$). It is interesting to note, however, that the tenuous disk emission remains important even for large effective radii.

An asymptotic limit is reached when the fraction of the flux emitted by the star goes to zero, in which case the relative weight of pseudo-photosphere {\it vs.} tenuous disk emission depends only on the density structure as
\begin{equation}
\lim_{\overline{R} \to \infty} \frac{F_{\lambda}^{\textrm{pphot}}}{F_{\lambda}}=\frac{2n-\beta-2}{2n-\beta},
\end{equation}
where $F_{\lambda}^{\textrm{pphot}}=B_{\lambda}(T_{\textrm{d}})\,A_{\textrm{pphot}}/d^2$. The tenuous disk emission fraction is simply the complement of this expression. For the the particular case where $n=3.5$ and $\beta=1.5$, the pseudo-photosphere contribution corresponds to $63.6\%$ of the total flux at large $\overline{R}$.


\psfrag{refrs}[c][][1.5]{$\overline{R}/R_{\star}$}

\begin{figure}
\begin{center}
\includegraphics[angle=0,scale=.8]{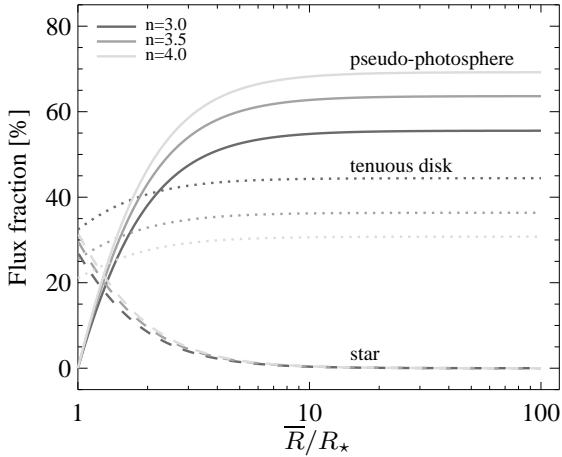}
\caption{Flux fraction of each emission component (star, pseudo-photosphere, tenuous disk, as indicated), as a function of effective radius. The models have $\beta=1.5$ and pole-on orientation. The disk size was assumed to be much larger than $\overline{R}$. \label{flux_fraction}}
\end{center}
\end{figure}


\subsection{Effect of disk truncation on the SED}
\label{trunc_sed}

In the situation where the the effective radius exceeds the disk physical size (Fig.~\ref{cases}c), the pseudo-photosphere no longer grows with wavelength. As a consequence, an abrupt change in the SED slope occurs (Eqs.~\ref{slp_dual_eq} and \ref{slp_trunc_eq}) at the wavelength $\lambda_{\textrm{trunc}}$, such that $\overline{R}(\lambda_{\textrm{trunc}})=R_{\textrm{d}}$. This situation is demonstrated in Fig.~\ref{sed_inflection}, that shows models with different disk radii. The IR excess no longer increases with wavelength, and the SED slope becomes identical to the stellar one. The inflection position therefore represents an important constraint to the disk size, as noted by \citet{waters1986}.


\subsection{Spectral slope dependence on disk parameters}
\label{slope_dependence}

Figure~\ref{slp_study} shows the dependence of the spectral slope with the disk density structure, based on the analytical model. The limiting case for large $\overline{R}$ (Eq.~\ref{slp_lim_eq}) is also shown.

The dependence with base density is stronger in the optical/near-IR, and becomes weak at longer wavelengths. Additionally, the value of $\alpha_{\textrm{IR}}$ is very sensitive to the parameter $n$, which causes a vertical shift of the entire curve. Finally, we can verify that for $\lambda\gtrsim 5\,{\rm {\mu}m}$, the limit for $\alpha_{\textrm{IR}}$ (Eq.~\ref{slp_lim_eq}) provides a very good approximation of the numerical differentiation of Eq.~(\ref{flux_dual}), with an accuracy of $5\%$.

This last result is particularly interesting, since it suggests that the IR spectral slope is a function of only $n$ and $\beta$, and depends very little on the disk density and inclination. However, since $n$ and $\beta$ appear as a linear combination ($2n-\beta$) in our formalism, it follows that $n$ and $\beta$ cannot be disentangled from SED observations only.


\subsection{Formation loci curves}
\label{loci}

\cite{carciofi2011} and \cite{rivinius2013} discuss the formation loci of different observables in terms of their enclosed flux fractions. The flux fraction $\mathscr{L}_{\lambda}$ at a given distance $R$ from the star is simply the integral of the specific intensity:
\begin{equation} \label{def_loci}
\mathscr{L}_{\lambda}(R)=\frac{2\pi\cos i}{F_{\lambda}^{\textrm{disk}}\,d^2}\,\int_{R_{\star}}^{R} I_{\lambda}(\varpi') \varpi'  d\varpi',
\end{equation}
where $F_{\textrm{disk}}$ the total disk flux. For the general case, integration of Eq.~(\ref{bright_dual_eq}) gives
\begin{equation}\label{loci_eq}
\mathscr{L}_{\lambda}(R) = \mathscr{L}_{0}
\begin{cases}
\frac{1}{2}[(R/R_{\star})^2-1] & \textrm{($R\leq \overline{R}$)}\\
c_1+c_2[1-(R/\overline{R})^{-2n+\beta+2}] & \textrm{($R> \overline{R}$)}\\
\end{cases} 
\end{equation}
where
\begin{align}
\mathscr{L}_{0}^{-1} &= \frac{1}{2}\left[\left(\overline{R}/R_{\star}\right)^2-1\right]+\frac{\overline{\tau}(\overline{R}/R_{\star})^2}{2n-\beta-2} \nonumber\\
                          & \quad\times\left[1-\left(R_{\textrm{d}}/\overline{R}\right)^{-2n+\beta+2}\right],
\end{align}
\begin{equation}
c_1=\mathscr{L}_{\lambda}(\overline{R}),
\end{equation}
and
\begin{equation}
c_2=\frac{\overline{\tau}(\overline{R}/R_{\star})^2}{2n-\beta-2}.
\end{equation}

Similar expressions can be written for a tenuous disk and a truncated pseudo-photosphere, based on Eqs.~(\ref{bright_diff_eq}) and (\ref{bright_trunc_eq}) respectively. Fig.~\ref{loci_curves} shows the brightness profiles and respective formation loci curves of a given model at three wavelengths, computed with both {\ttfamily HDUST} and the analytical model. We observe a very good agreement between the different methods. In the literature, the loci curves are often quoted in the context of estimating where in the disk a given observable (or, in our case, the continuum emission at a given wavelength) comes from. The pseudo-photosphere model (Eq.~\ref{loci_eq}) allows for an easy estimation of these curves without the need of complex RT numerical models.


\psfrag{refrs}[c][][1.5]{$\overline{R}/R_{\star}$}
\psfrag{Rstar}[c][][1.5]{$\,R_{\star}$}

\begin{figure}
\begin{center}
\includegraphics[angle=0,scale=0.65]{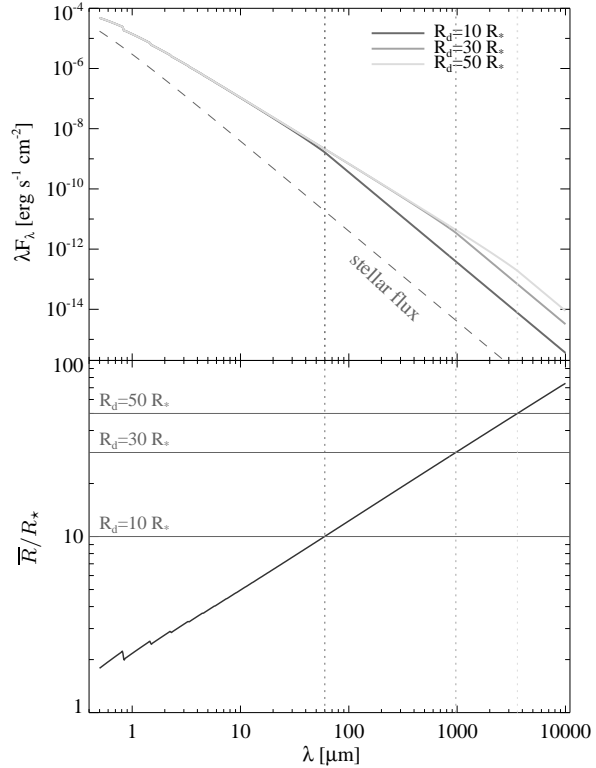}
\caption{{\it Top}: SED dependence on disk size. B1 spectral subtype, pole-on, $\rho_0=8.4\times 10^{-11}\,{\rm g\,cm^{-3}}$, $n=3.5$, at a distance of $50$~pc. {\it Bottom}: $\overline{R}$ dependence with wavelength for the same model. \label{sed_inflection}}
\end{center}
\end{figure}


\section{Application to IR data of Be stars}
\label{application}

As a first application of the pseudo-photosphere model, we study a sub-sample from the W87 list of objects, focusing on the objects with IRAS measurements at $12$, $25$ and $60\,{\rm{\mu}m}$. 

As mentioned in Sect.~\ref{introduction}, W87 interpreted IRAS observations of a large sample of Be stars based on an outflowing disk model that assumes a density radial profile
\begin{equation}
\rho(r)=\rho_0\left(\frac{r}{R_{\star}}\right)^{-n},
\end{equation}
where $r$ represents the distance to the star (in spherical coordinates). Also, W87 adopted a fixed opening angle of $\theta=15^{\circ}$, and restricted their modeling to the pole-on orientation. The last choice was based on the argument that for $i<70^{\circ}$, the disk inclination does not affect much the IR excess.

In order to compare both models, we adopt the same stellar parameters used by W87. The disk inclinations were taken from the literature, preferably from interferometric measurements (Table~\ref{par_emcee}). Since $\overline{R}$ is typically of the order of a few stellar radii at the mid-IR (Fig.~\ref{ref_lbd}), the IRAS fluxes probe the disk isothermal region \citep{carciofi2006}. Accordingly, we fixed the flaring parameter $\beta$ to $1.5$. We also chose $R_{\textrm{d}}=50\,R_{\star}$ ($\gg\overline{R}$ at $\lambda=60\,{\rm{\mu}m}$) and $\overline{\tau}=1.3$. Therefore, only $\rho_0$ and $n$ remain to be determined.

We fitted the color-corrected IR excesses of W87 using the {\ttfamily emcee} code \citep{foreman2013}, a Markov chain Monte Carlo (MCMC) implementation. Given a likelihood function of the parameters (in our case, $\propto \exp[-\chi^2]$), {\ttfamily emcee} provides samplings of the posterior probability. We started from uniform distributions for $\rho_0$ and $n$, and used 100 ``walkers'' (i.e., ensemble of random walking solutions) to explore the space of parameters. Each walker performed 20 steps in the burn-in phase, and 200 steps in the last phase, starting from the last burn-in chain state (for more details, see \citealt{foreman2013}).

\begin{figure}
\begin{center}
\includegraphics[angle=0,scale=.7]{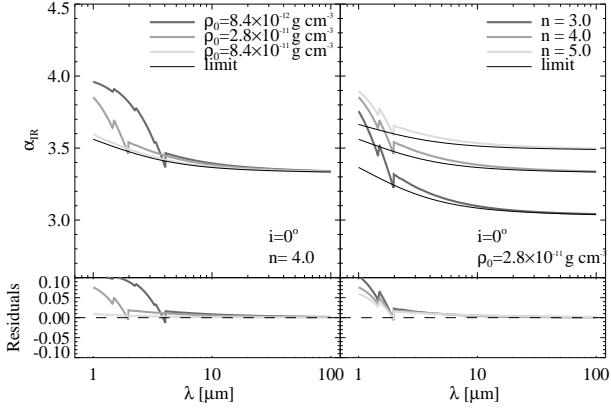}
\caption{Spectral slope calculated with the pseudo-photosphere model, for a representative set of parameters. {\it Left}: dependence with disk base density. {\it Right}: dependence with density radial slope (right panel). The limiting approximation for $\alpha_{\textrm{IR}}$ given by Eq.~(\ref{slp_lim_eq}) is represented by the black continuous lines. All the models have B1 spectral subtype and pole-on orientation. \label{slp_study}}
\end{center}
\end{figure}

The most likely parameters and the posterior probabilities for the object $\alpha$~Col are shown in Fig.~\ref{posterior_acol}, as an example of the typical typical results provided by {\ttfamily emcee} code (the plots for the remaining objects are available as online material). The ranges for both parameters were satisfactorily constrained despite the positive correlation between their values. This degeneracy occurs because $\overline{R}$ both increases with $\rho_0$ and decreases with $n$ (Eq.~\ref{ref_eq}). The list of fitted parameters is compared to the W87 results in Table~\ref{par_emcee}. The estimated uncertainties correspond to the confidence interval of $1\,\sigma$ of the posterior probability distributions.

Interestingly, the obtained results for $n$ and $\rho_0$ are quite similar to those found by W87, in spite of the VDD flaring.

\begin{figure}
\begin{center}
\includegraphics[angle=0,scale=.7]{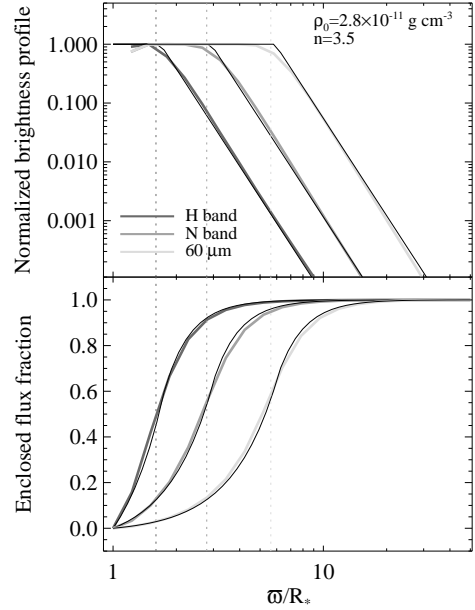}
\caption{Normalized brightness profiles at three spectral bands (top), and respective formation loci curves (bottom). The models have B1 subspectral type, pole-on orientation and disk parameters indicated. The {\ttfamily HDUST} results are represented in black, and the analytical models (gray lines) and the effective radii (vertical dashed lines) are superimposed. \label{loci_curves}}
\end{center}
\end{figure}


\psfrag{Alf Col}[c][][1]{$\alpha$~Col}

\begin{figure}
\begin{center}
\includegraphics[angle=0,scale=.8]{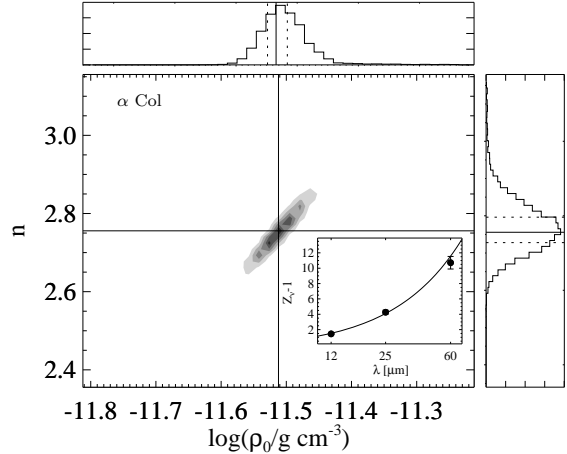}
\caption{Posterior probabilities for $\rho_0$ and $n$, computed for $\alpha$~Col, and the respective IR excess fit. The solid straight lines indicate the most likely values, while the dashed lines indicate the 1~$\sigma$ confidence interval. \label{posterior_acol}}
\end{center}
\end{figure}

\begin{table*}
\renewcommand{\footnoterule}{}  
\caption{Best-fit parameters obtained for $n$ and $\log(\rho_0)$, with $1\,\sigma$ confidence intervals estimated from the posterior probabilities computed with {\ttfamily emcee}. The results obtained by W87 are shown for comparison.}
\label{par_emcee}
\begin{center}
\begin{tabular}{llc | cc | ccc}
      \hline \hline
Name	&	HD	&	$i\,[\textrm{deg}]$	&	\multicolumn{2}{c|}{n}			&			\multicolumn{3}{c}{$\log(\rho_0/[{\rm g\,cm^{-3}}])$}			\\ 
	&		&		&	this work	&	W87	&	this work	&	\multicolumn{2}{c}{W87}			\\
	&		&		&		&		&		&	min	&	max	\\ \hline
$\gamma$~Cas	&	5394	&	40  $^{a}$	&	$3.17\substack{+0.10\\-0.06}$	&	3.25	&	$-10.81\substack{+0.07\\-0.04}$	&	-11.3	&	-10.6	\\
$\phi$~Per	&	10516	&	62 $^{a}$	&	$3.03\substack{+0.05\\-0.04}$	&	3.00	&	$-10.62\substack{+0.05\\-0.04}$	&	-11.1	&	-10.6	\\
$\psi$~Per	&	22192	&	57  $^{a}$	&	$2.58\substack{+0.03\\-0.02}$	&	2.50	&	$-11.23\substack{+0.02\\-0.02}$	&	-11.7	&	-11.4	\\
48~Per	&	25940	&	31 $^{a}$	&	$2.50\substack{+0.05\\-0.03}$	&	2.50	&	$-11.48\substack{+0.04\\-0.02}$	&	-11.8	&	-11.5	\\
$\zeta$~Tau	&	37202	&	73  $^{a}$	&	$3.26\substack{+0.07\\-0.05}$	&	3.25	&	$-10.59\substack{+0.05\\-0.06}$	&	-11.4	&	-10.9	\\
$\alpha$~Col	&	37795	&	35 $^{b}$	&	$2.76\substack{+0.04\\-0.03}$	&	2.75	&	$-11.51\substack{+0.02\\-0.01}$	&	-11.9	&	-11.7	\\
$\beta$~Mon	&	45725	&	67  $^{c}$	&	$3.88\substack{+0.03\\-0.12}$	&	3.00	&	$-10.13\substack{+0.03\\-0.10}$	&	-11.6	&	-10.9	\\
$\kappa$~CMa	&	50013	&	35  $^{b}$	&	$3.22\substack{+0.06\\-0.04}$	&	3.25	&	$-10.78\substack{+0.04\\-0.03}$	&	-11.3	&	-10.5	\\
$\beta$~CMi	&	58715	&	43  $^{d}$	&	$2.76\substack{+0.05\\-0.03}$	&	2.75	&	$-11.53\substack{+0.02\\-0.02}$	&	-12.0	&	-11.8	\\
$\omega$~Car	&	89080	&	65  $^{b}$	&	$2.72\substack{+0.04\\-0.03}$	&	2.75	&	$-11.54\substack{+0.03\\-0.02}$	&	-12.1	&	-11.9	\\
$\delta$~Cen	&	105435	&	35 $^{b}$	&	$2.61\substack{+0.04\\-0.03}$	&	2.50	&	$-11.22\substack{+0.04\\-0.03}$	&	-11.5	&	-11.2	\\
$\kappa$~Dra	&	109387	&	0  $^{e}$	&	$2.95\substack{+0.06\\-0.03}$	&	3.00	&	$-11.14\substack{+0.03\\-0.02}$	&	-11.5	&	-11.2	\\
$\chi$~Oph	&	148184	&	19  $^{e}$	&	$2.60\substack{+0.06\\-0.04}$	&	2.50	&	$-11.18\substack{+0.05\\-0.03}$	&	-11.6	&	-11.2	\\
$\alpha$~Ara	&	158427	&	45  $^{b}$	&	$3.07\substack{+0.07\\-0.05}$	&	3.00	&	$-10.65\substack{+0.04\\-0.04}$	&	-11.1	&	-10.6	\\
66~Oph	&	164284	&	47  $^{e}$	&	$2.54\substack{+0.04\\-0.03}$	&	2.50	&	$-11.16\substack{+0.03\\-0.02}$	&	-11.5	&	-11.1	\\
\hline \hline
\end{tabular}
\end{center}
(a) \citet{quirrenbach1997}; (b) \citet{meilland2012}; (c) \citet{fremat2005}; (d) \citet{klement2015}; and (e) \citet{touhami2013}.
\end{table*}


\subsection{Mass decretion rates}
\label{decretion}

Under the assumption of an isothermal disk and purely Keplerian circular orbits, the VDD analytical solution for the surface density is given by (e.g., \citealt{bjorkman2005})
\begin{equation}
\Sigma(\varpi)=\frac{\dot{M}\,V_{\textrm{crit}}\,R_{\star}^{1/2}}{3\pi \alpha c_s^2\,\varpi^{3/2}}\left[\left(R_0/\varpi\right)^{1/2}-1 \right],
\end{equation}
where $\alpha$ represents the viscosity parameter introduced by \citet{shakura1973}, and $R_0$ is an arbitrary integration constant associated to the disk size. It is related to the volume density as 
\begin{equation}\label{sig_rho}
\Sigma(\varpi)=\int_{-\infty}^{+\infty}\rho(\varpi,z)\,dz=\sqrt{2\pi}\,H(\varpi)\,\rho(\varpi,0).
\end{equation}
By evaluating this equation at $R_{\star}$ and rearranging its terms, we can write the expression for the mass decretion rate:
\begin{equation}\label{mdot_def}
\dot{M}=\frac{3\pi\sqrt{2\pi}\alpha\,R_{\star}^2\, c_s^3\,\rho_0}{V_{\textrm{crit}}^{2} \left[\left(R_0/R_{\star}\right)^{1/2}-1\right]}.
\end{equation}

In order to estimate a typical range for $\dot{M}$, we consider a disk that is not perturbed by external tidal forces, which is the case for Be either isolated stars or those belonging to a well detached binary system. The $R_0$ can then be approximated by the isothermal critical radius \citep{krticka2011,okazaki2001}
\begin{equation}
\frac{R_0}{R_{\star}}=\frac{R_{\textrm{c}}}{R_{\star}}=\frac{3}{10}\left(\frac{V_{\textrm{crit}}}{c_s}\right)^2,
\end{equation}
which represents the transition point between the subsonic inner part ($v_{\varpi}\ll c_s$) and transonic outer part ($v_{\varpi}\gtrsim c_s$) of the disk. The derived values for $R_{\textrm{c}}$ range from $320$ to $650\,R_{\star}$, depending on the adopted stellar parameters. Since $\alpha$ is unknown, upper and lower limits for $\dot{M}$ were estimated by choosing $\alpha=1$ and $\alpha=0.1$, respectively.

The mass decretion rates thus determined are listed in Table~\ref{mass_loss_fit} and plotted in Fig.~\ref{lum_mdot} along with the results of W87. The values of $\dot{M}$ estimated from the VDD model are two to three orders of magnitude smaller than the results of W87, depending on the value of $\alpha$. Our results are compatible with the mean mechanical mass loss values of \citet{granada2013}, predicted for critically rotating stellar models.

\begin{table}
  \caption{Mass decretion rates for non-perturbed VDD, and previous values found by W87.}
  \label{mass_loss_fit}
  \begin{center}
    \begin{tabular}{l | cc |c}
      \hline \hline
Name	&	\multicolumn{3}{c}{$\log(\dot{M}/[{\rm M_{\odot}\, yr^{-1}}])$}					\\ 
	&	\multicolumn{2}{c|}{This work}			&	W87	\\
	&	$\alpha=0.1$	&	$\alpha=1.0$	&		\\ \hline
$\gamma$~Cas	&	-10.1	&	-9.1	&	-6.9	\\
$\phi$~Per	&	-10.4	&	-9.4	&	-7.3	\\
$\psi$~Per	&	-11.3	&	-10.3	&	-7.9	\\
48~Per	&	-11.4	&	-10.4	&	-7.9	\\
$\zeta$~Tau	&	-10.6	&	-9.6	&	-7.6	\\
$\alpha$~Col	&	-12.1	&	-11.1	&	-8.8	\\
$\beta$~Mon	&	-10.7	&	-9.7	&	-8.1	\\
$\kappa$~CMa	&	-10.6	&	-9.6	&	-7.3	\\
$\beta$~CMi	&	-12.7	&	-11.7	&	-9.0	\\
$\omega$~Car	&	-11.9	&	-10.9	&	-8.6	\\
$\delta$~Cen	&	-11.0	&	-10.0	&	-7.6	\\
$\kappa$~Dra	&	-11.5	&	-10.5	&	-7.9	\\
$\chi$~Oph	&	-11.0	&	-10.0	&	-7.8	\\
$\alpha$~Ara	&	-11.0	&	-10.0	&	-7.6	\\
66~Oph	&	-11.5	&	-10.5	&	-7.7	\\
\hline \hline
\end{tabular}
\end{center}
\end{table}


\psfrag{logLstr}[c][][1.5]{$\log(L/{\rm L_{\odot}})$}
\psfrag{logMdotstr}[c][][1.5]{$\log(\dot{M}/[{\rm M_{\odot}\,yr^{-1}}])$}

\begin{figure}
\begin{center}
\includegraphics[angle=0,scale=.8]{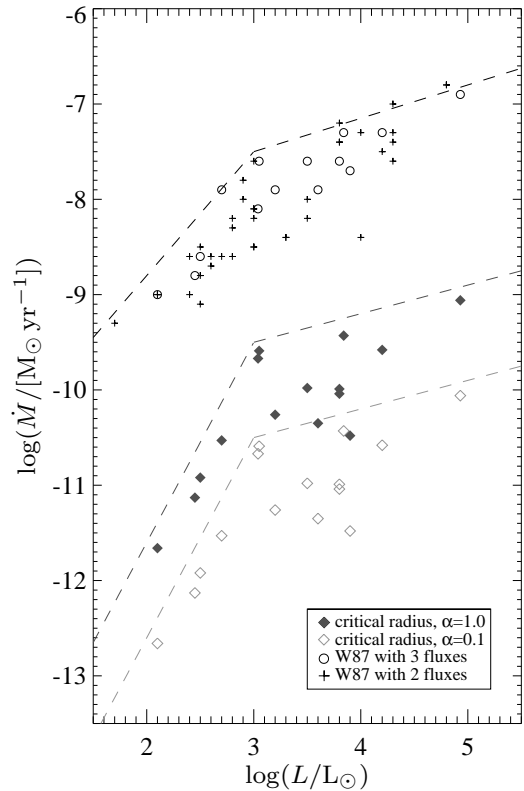}
\caption{Mass loss-luminosity diagram for Be stars. The symbols in black represent the results obtained by W87, while the black dashed line represents the upper limit suggested by the same authors. The gray dashed lines indicate the upper limits found by our study (Eq.~\ref{lum_lim}). \label{lum_mdot}}
\end{center}
\end{figure}

The large discrepancy with the W87 results must be examined in more detail. Firstly, the definition of mass decretion rate must be recalled. In the VDD formalism, the mass decretion rate is the mass that flows outward through the disk per unit time, and is related the outflow speed ($v_{\varpi}$) and the surface density by the mass conservation relation:
\begin{equation}
\dot{M}=2\pi\varpi v_{\varpi} \Sigma.
\end{equation}
Physically, it corresponds to the net mass that is lost by the star through the disk. The decretion requires an outward momentum flux, and the angular momentum is taken from part of the injected mass. Therefore, a large fraction of the mass that is ejected from the star and fed into the disk falls back (reacretes) to the photosphere. For instance, \citet{carciofi2012} determined the mass injection rate of the outburst phase of 28~CMa, a B2IV-Ve star, to be $(3.5\pm1.3)\times 10^{-8}\,{\rm M_{\odot}\,yr^{-1}}$. However, the mass decretion rate in that case is about $10^{-3}$ smaller, and thus compatible with the results of Fig.~\ref{lum_mdot}.

The W87 outflow model was inspired in the CAK model for radiatively driven winds \citep{castor1975}. Although it predicts more material distributed above the equatorial plane than the VDD model, W87 disk mass is only a few times larger than the VDD one ($\simeq 4$~times for a B1 star with $R_{\textrm{d}}/R_{\textrm{e}}=50$ and $n=3.5$). The large differences of $\dot{M}$ between W87 and this work comes mostly by largely different outflow speeds in each model. While W87 adopted outflow speeds of the order of $5\,{\rm km\,s^{-1}}$ at the disk base, the VDD model predicts much smaller $v_{\varpi}$ values. For $\alpha=1$, $v_{\varpi}/c_s\sim 10^{-3}$ at the stellar equator \citep{krticka2011}, and therefore $v_{\varpi}\sim 10^{-2}\,{\rm km\, s^{-1}}$. The absence of large scale radial motions in Be disk was observationally established by spectroscopy \citep{hanuschik1995}, spectro-interferometry \citep{meilland2012} and spectro-astrometry \citep{wheelwright2012}, and is one of the most important facts in support of the VDD model \citep{rivinius2013}. We conclude that the differences in $\dot{M}$ between this works and W87 is due to the $\sim$two orders of magnitude difference in the radial outflow speeds.

Fig.~\ref{lum_mdot} shows that, despite the aforementioned large differences in the absolute value of $\dot{M}$, both models agree qualitatively in that later type stars have much smaller $\dot{M}$ than early types ones. This is an important result, which may be in close connection to the Be phenomenon itself. For reference, the following formulae express the upper limits $\dot{M}$ shown in Fig.~\ref{lum_mdot}, estimated for the $\alpha=1$ models:
\begin{equation}\label{lum_lim}
\log(\dot{M})=
\begin{cases}
-15.8+2.1\,\log(L/L_{\odot}), & \textrm{$\log(L/L_{\odot})\leq 3$}\\
-10.4+0.3\,\log(L/L_{\odot}), & \textrm{$\log(L/L_{\odot})> 3$}.\\
\end{cases}
\end{equation}


\section{Conclusions}

The continuum brightness profile of gaseous disks is investigated. The viscous decretion disk model of Be star disks was used as an example, but the concepts here presented can be easily extended to other types of objects, such as Herbig~Ae/Be stars. Numerical results indicate that the profile can be interpreted in terms of two components: an inner, optically thick part, dubbed the pseudo-photosphere, a term borrowed from the LBVs and supernovae literature, and an outer, tenuous part. 

Based on this realization, a semi-analytical model was developed that is capable of satisfactorily describing the size of the pseudo-photosphere and the disk SED with an accuracy better than 20\% for the flux level and 5\% for the spectral slope, when compared to numerical results. The model is said to be semi-analytical because one of its main parameters, namely the optical depth at the line of sight direction ($\overline{\tau}$), defining the size of the pseudo-photosphere ($\overline{R}$), was empirically measured to be around $1.3$ using the numerical results.

This model allowed identifying important properties of the emergent flux, in relation to fundamental parameters of the system. For instance, the IR spectral slope is shown to be dependent only on the radial slope of the density, $n$, and the disk flaring parameter, $\beta$, while being rather insensitive to $\rho_0$, $i$, and even the stellar parameters. This result is particularly interesting, since it provides an easy and direct way to extract information about the physical structure of the disk from IR spectra or colors, without prior knowledge about the central star.

As a first application of the model, a sample of 15 Be stars were studied and compared to previous results by \citet{waters1987}, that used an outflowing model to describe the disk structure. This allowed a first comparison between the VDD model and the widely quoted results by \citeauthor{waters1987}. It was found that, while the density parameters of the disk are generally in good agreement, the mass decretion rates were two to three orders of magnitude smaller than previously estimated. This result has important consequences for evolutionary models of Be stars as class of fast-rotating stars that shed angular momentum through the viscous disk, and to the Be phenomenon itself.


\section*{Acknowledgments}

We thank the referee, prof. Carol Jones, for her useful comments. This work made use of the computing facilities of the Laboratory of Astroinformatics (IAG/USP, NAT/Unicsul), whose purchase was made possible by the Brazilian agency FAPESP (grant 2009/54006-4) and the INCT-A.  R.~G.~V. acknowledges the support from FAPESP (grant 2012/20364-4). A.~C.~C acknowledges support from CNPq (grant 307076/2012-1).



\appendix

\section{Adopted gaunt factors and fitting formulae}\label{gaunt_factor}

In the present work, we adopted the bound-free and free-free gaunt factors of \citet{hummer1988} and \citet{storey1991}, where the gas material is assumed to be exclusively composed by hydrogen. In order to provide approximate expressions for such quantity, we propose the following fitting formulae:
\begin{equation}\label{ff_fit}
g(\lambda,T)\simeq \exp\left[G_0(T)+G_1(T)\,\ln\lambda+G_2(T)\,(\ln\lambda)^2  \right],
\end{equation}
\begin{equation}\label{bf_fit}
b(\lambda,T)\simeq \exp\left[B_0(T)+B_1(T)\,\ln\lambda+B_2(T)\,(\ln\lambda)^2  \right].
\end{equation}
One of the advantages of the adopted expressions is the simplification of the gaunt factors spectral slope to
\begin{equation}\label{u_fit}
u(\lambda,T) = \frac{1}{g+b}[(G_1+2 G_2\ln\lambda)g+(B_1+2 B_2\ln\lambda)b].
\end{equation}
The numerical gaunt factors were fitted in the range between $1\,{\rm \mu m}$ and $1\,{\rm mm}$, at six temperatures ranging from $5,000$~K to $20,000$~K. The fitted coefficients are presented in Table~\ref{gf_fit_par}. The numerical and approximated gaunt factors are shown in Fig.~\ref{gf_fit}. Our fit expressions reproduce the gaunt factors within an accuracy of $3$\%, in the spectral range between $10\,{\rm \mu m}$ and $1\,{\rm mm}$. For shorter wavelengths, the bound-free gaunt factor is dominant, and its slope discontinuities cannot be fitted by our simple relations. However, the approximation at the near-IR is increasingly improved for higher temperatures.

\begin{table}
  \caption[Gaunt factors fit parameters.]
{Gaunt factors fit parameters.}
{\scriptsize
  \label{gf_fit_par}
  \begin{center}
    \begin{tabular}{lcccccc}
      \hline \hline
$\log(T/{\rm K})$	&	$G_0$	&	$G_1$	&	$G_2$	&	$B_0$	&	$B_1$	&	$B_2$	\\	\hline
      $3.70$ &      $0.0952$ &      $0.0215$ &      $0.0145$ &        $2.2125$ &       $-1.5290$ &      $0.0563$ \\
      $3.82$ &       $0.1001$ &      $0.0421$ &      $0.0130$ &        $1.6304$ &       $-1.3884$ &      $0.0413$ \\
      $3.94$ &       $0.1097$ &      $0.0639$ &      $0.0111$ &        $1.1316$ &       $-1.2866$ &      $0.0305$ \\
      $4.06$ &       $0.1250$ &      $0.0858$ &     $0.0090$ &       $0.6927$ &       $-1.2128$ &      $0.0226$ \\
      $4.18$ &       $0.1470$ &       $0.1071$ &     $0.0068$ &       $0.2964$ &       $-1.1585$ &      $0.0169$ \\
      $4.30$ &       $0.1761$ &       $0.1269$ &     $0.0046$ &     $-0.0690$ &       $-1.1185$ &      $0.0126$ \\
\hline
\end{tabular}
\end{center}
}
\end{table}


\begin{figure}
\begin{center}
\includegraphics[angle=0,scale=.8]{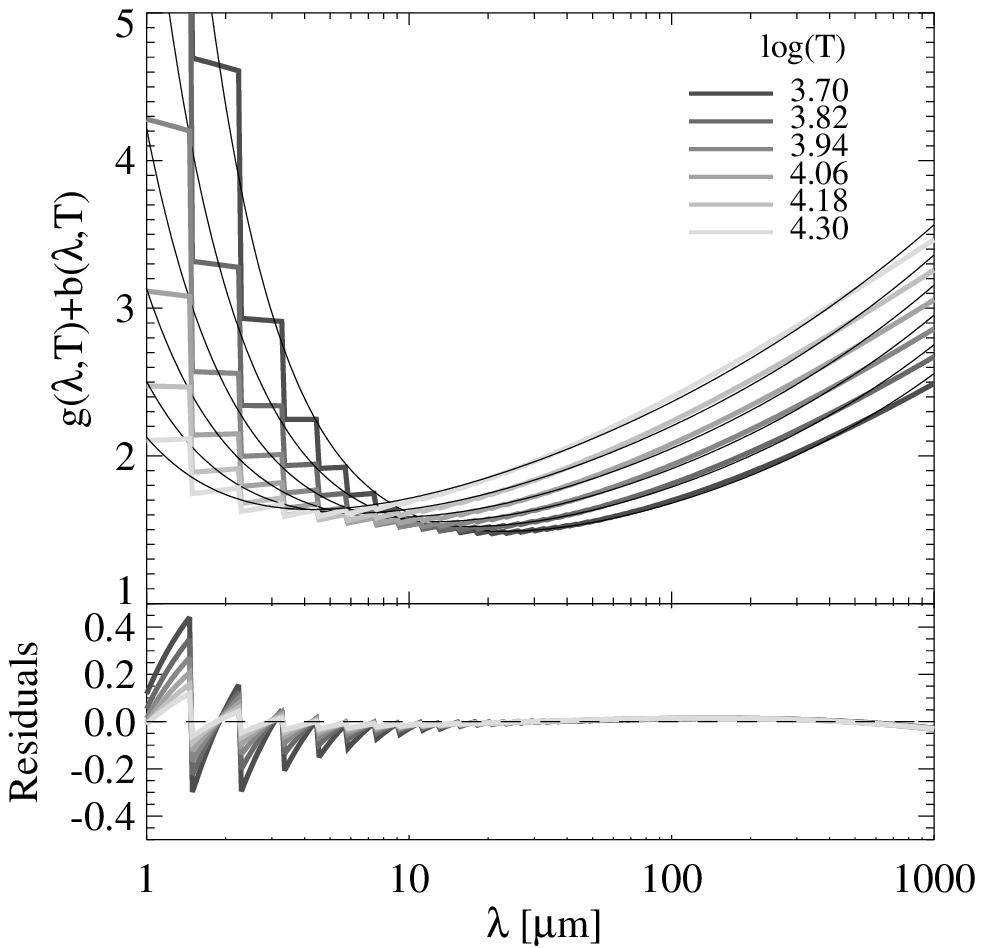}
\caption{Gaunt factors of \citet{hummer1988} and \citet{storey1991} for several temperatures, as indicated, compared to the fitting expressions given by Eqs.~\ref{ff_fit} and \ref{bf_fit}. \label{gf_fit}}
\end{center}
\end{figure}

\section{Black body correction}
\label{bb_corr_factor}

Approximating the stellar flux by a black body systematically overestimates the IR flux, when compared to more realistic atmospheric stellar models (Fig.~\ref{kur_bb_comparison}). In order to compensate for this effect, we computed the ratio between Kurucz model atmosphere fluxes (\citealt{castelli2003}) and the corresponding black body emission at $10\,{\rm\mu{m}}$
\begin{equation}
f^{\star}=\left. \frac{F_{\lambda}^{\textrm{K}}(T_{\textrm{eff}},\log g)}{F_{\lambda}^{\textrm{bb}}(T_{\textrm{eff}})}\right|_{\lambda=10\,{\rm\mu m}}.
\end{equation}
Fig.~\ref{bb_correction} shows the defined ratio as a function of the stellar effective temperature. It can be used as a black body correction factor, to improve the accuracy of our semi-analytical model, and can be approximated by
\begin{equation}
f^{\star}\simeq 1.015-0.301\left(\frac{T_{\textrm{eff}}}{10^4\,{\rm K}}\right)+0.064\left(\frac{T_{\textrm{eff}}}{10^4\,{\rm K}}\right)^2.
\end{equation}


\begin{figure}
\begin{center}
\includegraphics[angle=0,scale=.8]{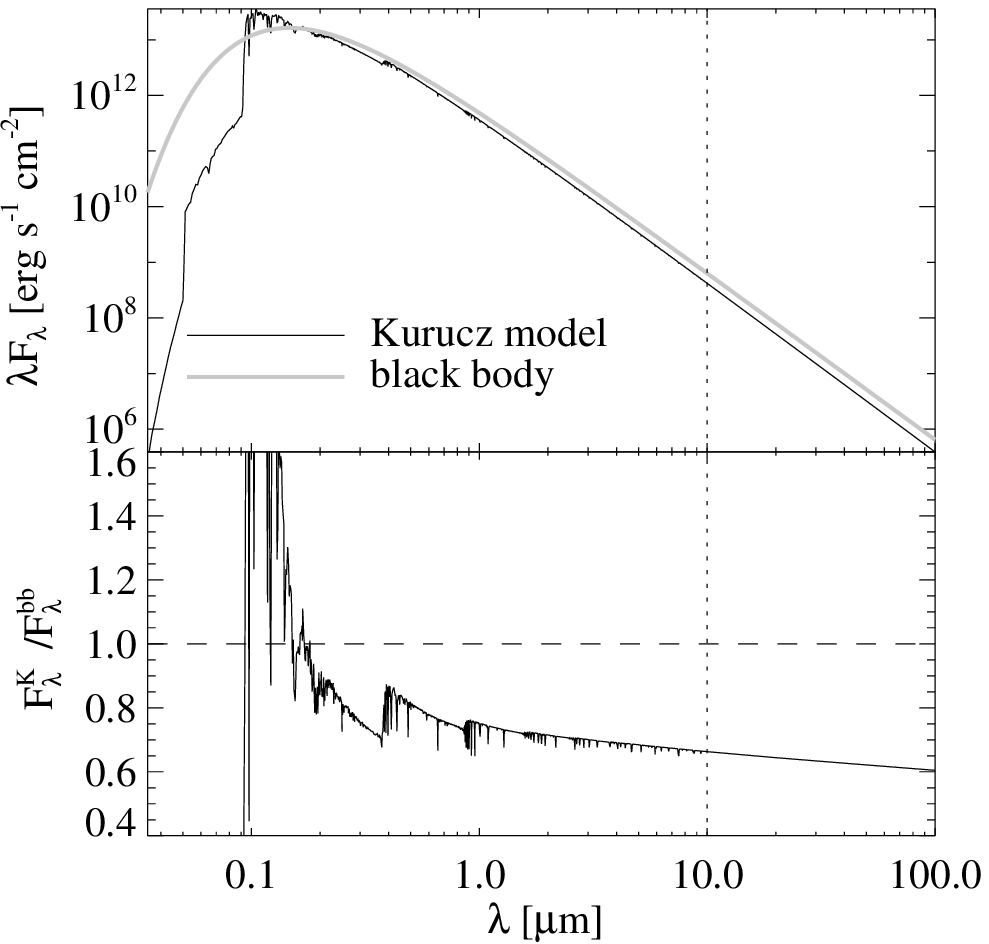}
\caption{Comparison between Kurucz model ($\log g=4$ fixed) and black body fluxes at the stellar surface, for $T_{\textrm{eff}}=26,000\,{\rm K}$. The position $\lambda=10\,{\rm{\mu}m}$ is indicated by the vertical dotted line. \label{kur_bb_comparison}}
\end{center}
\end{figure}


\psfrag{fs}[c][][1.5]{$f^{\star}$}

\begin{figure}
\begin{center}
\includegraphics[angle=0,scale=.8]{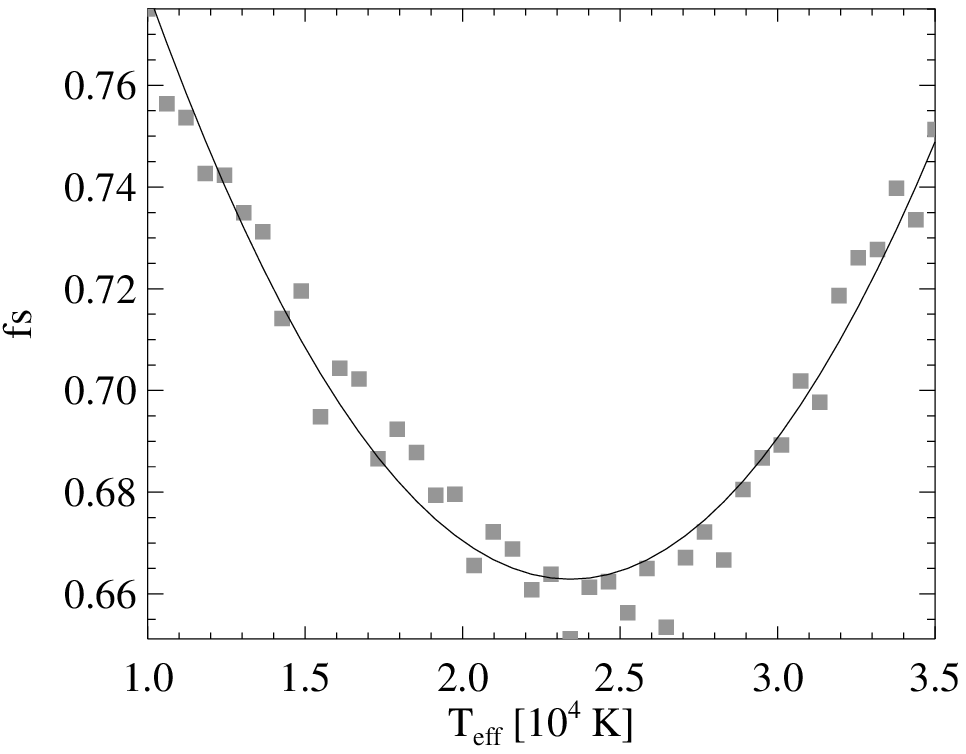}
\caption{Black body correction factor, as a function of $T_{\textrm{eff}}$. A second degree polynomial fit is superimposed to the computed values. \label{bb_correction}}
\end{center}
\end{figure}

\bsp

\label{lastpage}

\end{document}